\documentclass[twocolumn,prl,aps]{revtex4}
\usepackage{graphicx}

\begin{document}

\title{Long distance quantum teleportation in a quantum relay configuration}

\author{H. de Riedmatten, I. Marcikic, W. Tittel, H. Zbinden, D. Collins and N.
Gisin}

\affiliation{Group of Applied Physics, University of Geneva,
Switzerland\\}
\begin{abstract}
A long distance quantum teleportation experiment with a
fiber-delayed Bell State Measurement (BSM) is reported. The source
creating the qubits to be teleported and the source creating the
necessary entangled state are connected to the beam splitter
realizing the BSM by two 2\,km long optical fibers. In addition,
the teleported qubits are analyzed after 2,2 km of optical fiber,
in another lab separated by 55\,m. Time bin qubits carried by
photons at 1310\,nm are teleported onto photons at 1550\,nm. The
fidelity is of $77\,\%$, above the maximal value obtainable
without entanglement. This is the first realization of an
elementary quantum relay over significant distances, which will
allow an increase in the range of quantum communication and
quantum key distribution.
\end{abstract}

\maketitle Quantum teleportation (QT) is the transmission of the
quantum state of a particle to another distant one, without the
transmission of the particle itself. The QT channel consists of
non classical Einstein-Podolski-Rosen (EPR) correlations between
two particles supplemented by some bits of classical information.
Since the theoretical proposal in 1993 \cite{Bennett93}, several
experiments have demonstrated the principle of QT using different
types of discrete and continuous variables
\cite{Bouwmeester97,Boschi98,Kim2001,KimbleTelep,Lombardi02,Marcikic03}.
QT could find applications in the context of quantum
communication, where the goal is to distribute quantum states over
large distances, e.g for quantum key distribution \cite{rmp} or
for future quantum networks \cite{cirac97}. The simplest way to
send a quantum state from a sender to a receiver, usually called
Alice and Bob, is to send directly the particle carrying the
state, e.g using an optical fiber. However, as detectors are noisy
and fibers lossy, the signal to noise ratio (and thus the
fidelity) decreases with distance, and the maximal distance for a
given fidelity is thus limited. Quantum repeaters \cite{briegel98}
that rely on QT, entanglement purification\cite{pan03} and quantum
memories have been proposed to overcome this problem. However, a
full quantum repeater is not realizable with present technology,
due to the need for a quantum memory. Nevertheless, QT could still
be useful in this context, even without quantum memory, by
implementing a so called quantum relay \cite{rmp,relay,Collins03}.

\begin{figure}[h]
\includegraphics[width=7cm]{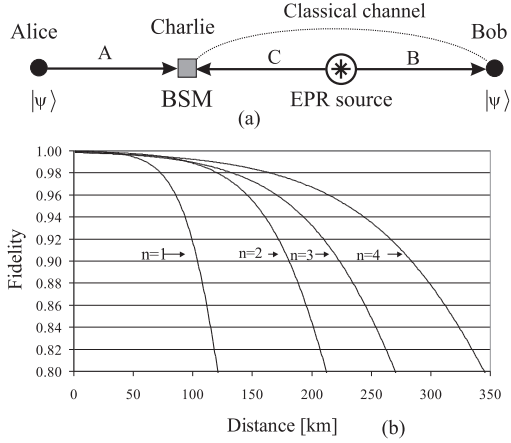}
\caption{(a)Quantum teleportation as a quantum relay. (b) Fidelity
of the transmitted quantum state as a function of the distance for
different configurations. Direct transmission (n=1), with an EPR
source in the middle (n=2), teleportation (n=3), and entanglement
swapping (n=4). We assume that the fidelity is only affected by
the detectors noise. The curves are plotted for a realistic dark
count probability $D=10^{-4}$per ns and a fiber attenuation of
$0.25db/km$. \label{relay}}
\end{figure}

The basic idea is the following. Suppose that the qubit sent by
Alice is teleported to Bob using the entangled state created by an
EPR source, as in Fig \ref{relay} a.  To perform the
teleportation, Charlie makes a joint measurement, called a Bell
State Measurement (BSM) between Alice's photon and one half of the
EPR pair, which projects Bob's photon into the state of Alice's
photon, modulo a unitary transformation. A successful BSM implies
in particular that a photon has left the EPR source towards Bob.
This means that although the logical qubit travels an overall
distance $l$ from Alice to Bob, the effective distance covered by
the photon to be detected by Bob is reduced to around $l/3$. This
provides an increase in the maximal distance for a given fidelity.
Fig \ref{relay}b shows the calculated qubit fidelity as a function
of the distance, for different configurations (see
\cite{Collins03} for the details of the calculations). The price
to pay is a reduction of the count rate. A first experiment in
this direction has been reported recently, where the teleported
qubit was analyzed after 2 km of optical fibers \cite{Marcikic03}.
But this is not enough, as the source creating the qubit and the
EPR source should be distant from each other. In this paper, we
present the first realization of an elementary quantum relay
 by implementing a long distance
QT experiment with a fiber-delayed BSM in order to simulate
distant sources.

Beyond experimental difficulties encountered in short distance QT
experiments, the addition of distance implies two additional
challenges. First, entanglement must be distributed over large
distance with a high quality. This is overcome by taking profit of
the robustness of time-bin entanglement when transmitted in
optical fibers \cite{ThewRobust}. Moreover, the use of standard
telecommunication fibers imposes the wavelength of the carrier
photons, in order to minimize losses.
Second, to achieve a successful BSM, the two photon involved must
be indistinguishable, which is much more difficult to preserve
when the photons travel long distances in two optical fibers. In
the following, we will first review the basic principles of
quantum teleportation with time-bin qubits and then present the
experiment, focussing mainly on the indistinguishability of the
photons involved in the BSM.

The procedure of quantum teleportation using time-bin qubits is
the following. Alice prepares a time-bin qubit, that is a photon
in a coherent superposition of two time bins, by passing a single
photon through a unbalanced interferometer with path length
difference $\Delta \tau$. The time bin qubit can be written as: $
\left|
\psi\right\rangle_A=a_0\left|1,0\right\rangle_A+a_1e^{i\alpha}\left|0,1\right\rangle_A
, $where$\left|1,0\right\rangle_A$ represents the first time-bin
(i.e a photon having passed through the short arm of the
interferometer), $\left|0,1\right\rangle_A$ the second time bin
(i.e a photon having passed through the long arm), $\alpha$ a
relative phase and $a_0^2+a_1^2=1$. Time-bin qubits can be
conveniently represented on a Poincar\'e sphere (see Fig.2b).
Alice sends the qubit to Charlie, who shares with Bob a classical
communication channel and a pair of time-bin entangled
qubits\cite{marcikic02} in the state: $ \left| \phi
^+\right\rangle_{BC}=\frac{1}{\sqrt{2}}(\left| 1,0
\right\rangle_C\left|1,0\right\rangle_B+ \left| 0,1
\right\rangle_C\left|0,1\right\rangle_B) \label{phi} $. Charlie
now performs the BSM i.e. he projects photons A and C onto one of
the four Bell States, defined by : $
 \left| \phi
^{\pm}\right\rangle=\frac{1}{\sqrt{2}}(\left| 1,0
\right\rangle\left|1,0\right\rangle\pm \left| 0,1
\right\rangle\left|0,1\right\rangle)$ and $
 \left| \psi
^{\pm}\right\rangle=\frac{1}{\sqrt{2}}(\left| 1,0
\right\rangle\left|0,1\right\rangle\pm \left| 0,1
\right\rangle\left|1,0\right\rangle) $. Depending on the result of
the BSM (communicated by Charlie with  two classical bits), Bob
can now apply the appropriate transformation (i.e, bit flip and/or
phase flip or identity) to recover the initial state. We perform
an interferometric BSM, by mixing the two photons on a beam
splitter (one per input mode) that allows one in principle to
discriminate two out of the four Bell states \cite{Lutkenhaus}. We
select only projections onto the $\left| \psi ^{-}\right\rangle$
singlet state. It can be shown that the detection of one photon in
each output mode with a time difference $\Delta \tau$ realizes
this projection. In this case, photon B is projected onto the
state:
\begin {equation}
\left| \psi \right\rangle_B = a_0\left|0,1
\right\rangle_{B}-a_1e^{i\alpha} \left|1,0
\right\rangle_{B}=i\sigma_y\left| \psi \right\rangle_A
\label{teleport}
\end{equation}
where $\sigma_y$ is a Pauli matrix.

\begin{figure}[h]
\includegraphics[width=7.5cm]{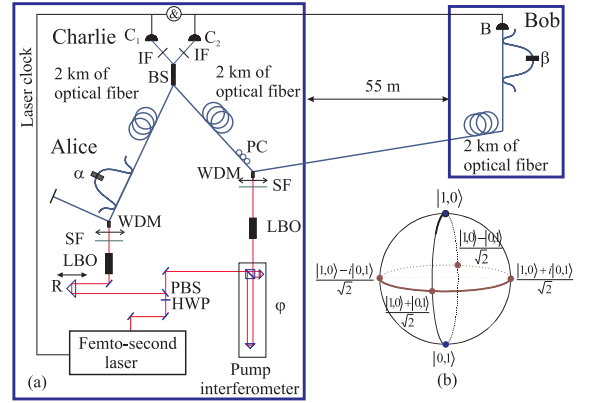}
\caption{(a)Long distance quantum teleportation experiment, with
2km of optical fibers inserted between each source and the BSM.
(b) Poincar\'e sphere representing the qubit states}\label{setup6}
\end{figure}

A schematic of the experiment is presented in Fig. \ref{setup6}a
(see also \cite{Marcikic03} for a more detailed description).
Ultra-short pulses from a mode-locked Ti-Sapphire laser ($\Delta
t=150fs, \lambda= 710nm, f_{rep}=75 MHz$) are split by a variable
beam-splitter made of a half-wave plate and a polarizing beam
splitter (HWP+PBS). The transmitted beam is used to create
entangled time-bin qubits in the state $ \left| \phi
^+\right\rangle_{BC}$, by passing it first through a bulk optical
Michelson interferometer (pump interferometer) with path-length
difference $\Delta\tau=1.2ns$, and then through a non linear
Lithium triborate (LBO) crystal, where entangled non degenerate
collinear time-bin qubits at telecom wavelength (1310 and 1550 nm)
are created. The optical path length difference of this
interferometer defines a reference, therefore the phase $\varphi$
is taken to be zero. The pump beam is removed with a Silicon
filter (SF) and the created photons are coupled into an optical
fiber and separated using a wavelength division multiplexer (WDM).
Photon C at 1310 nm is sent to Charlie and its twin photon B at
1550nm to Bob (see Fig 1a). \\The beam reflected at the variable
coupler is used to create the qubits to be teleported. To do so,
the 1310 nm photon (A) from a pair produced in a similar LBO
crystal passes through a fiber Michelson interferometer with
relative phase $\alpha$ generating a superposition between two
time-bins. To create the two time-bin states
$\left|1,0\right\rangle_{A}$ and $\left|0,1 \right\rangle_{A}$,
fibers of appropriate length are employed instead of the
interferometer. The retroreflector R is used to adjust the qubit
arrival time at the BSM. The two 1310 nm photons (A and C) travel
to Charlie, who performs the BSM with a 50/50 fiber beam splitter
(BS), through 2 km of standard optical fibers. In order to avoid
spurious coincidences that would limit the teleportation fidelity,
one has to post select only events where one pair is created in
each crystal. This is done by decreasing the probability of
creating a pair in the EPR source, relative to the qubit source by
a factor of seven using the variable coupler \cite{Marcikic03}.
\\
Bob is located in another lab, separated by 55m, and connected by
2,2 km of dispersion-shifted optical fiber. The qubit is analyzed
with a fiber interferometer similar to the one used for the
creation, but adapted for the wavelength of 1550nm or by a fiber (see below). \\
All photons are detected with avalanche photodiodes (APD).
Detector $C_1$ is a passively quenched Ge APD (quantum efficieny
$\eta =10\%$, dark count rate dc=40kHz), and detectors $C_2$ and B
are Peltier cooled InGaAs APDs operating in gated mode
($\eta=30\%$, dc=$10^{-4}$ per ns)\cite{stucki01}. To reduce the
dark count rate, the trigger signal for the latter is only given
by a coincidence between detector $C_1$ and the laser clock
($t_0$). The BSM (coincidence $C_1+C_2+ t_0)$ is realized using a
fast (i.e $<$ 1ns) coincidence electronics. To verify the
teleportation process, four fold coincidences ($C_1+C_2+B+t_0$)
are monitored with a Time-to Digital Converter (TDC), where the
start is given by a successful BSM and the stop by Bob's detector
B.

\begin{figure}[h!]
\includegraphics[width=7cm]{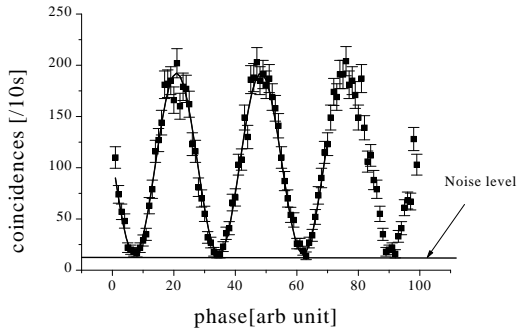}
\caption{Two photon interference after two times 2 km of optical
fibers. The solid line is a sinusoidal fit from which we can infer
a net visibility of $96 \pm 1 \%$ } \label{testbell6km}
\end{figure}

To verify that the entanglement of the EPR pair is preserved after
the transmission over 2x2 km of optical fibers, we perform a
Franson-type two photon interference experiment with the same 3
interferometers used for the teleportation. The method is
described in details in \cite{marcikic02}. We observe interference
fringes with a high net visibility of $96 \pm 1 \%$ (see fig
\ref{testbell6km}), showing that there is no degradation of the
entanglement during the transmission. This measurement also allows
us to insure that the three interferometers have the same optical
path length difference, which is also required for the
teleportation experiment.

Let us now analyze in more detail the indistinguishability
criteria for the BSM: photons A and C must be described by
identical spatial, spectral, polarization and temporal modes.
Spatial and spectral indistinguishability is ensured by using a
fiber beam splitter and identical interference filters (IF) for
both photons respectively. A polarization controller (PC) inserted
before the BSM beam splitter allows us to equalize the
polarization for the two photons. For temporal
indistinguishability, two points must be considered. First, the
photons must be created at well defined times, which means that
their coherence time must be larger than the pump pulse duration
\cite{zukowski95}. This is achieved by using ultrashort pulses
(150\,fs) and narrow IF (10\,nm) to increase the coherence time of
the photons to about 250\,fs. Second, the two photons must arrive
at the same time at the BSM beam splitter, within their coherence
time. This implies in particular that the length difference of the
two fibers must remain constant within a few tenth of microns for
several hours.
\begin{figure}[h!]
\includegraphics[width=7cm]{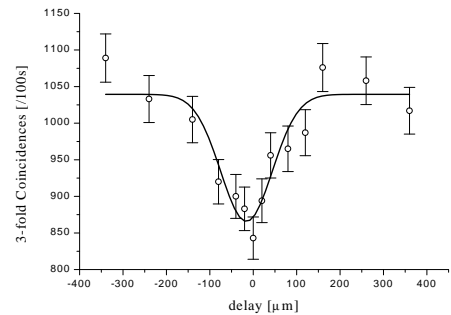}
\caption{HOM interference after 2x2 km of optical fibers. See text
for details} \label{dip3x2km}
\end{figure}
Thermally induced fiber length variations are of the order of 8
mm/K for a two km long fiber. The temperature of the two fiber
spools should thus be constant within the mK range. Even in a well
controlled laboratory environment, this proved to be a very
stringent condition. We finally placed both fibers together on the
same spool. In this case, local temperature variations act in the
same way on the two fibers. With this method, we obtained length
difference variations of around $3 \mu m/h$, an acceptable value
for a teleportation experiment. Finally, the two fibres must have
the same chromatic dispersion in order to avoid "which path"
information. Indeed, contrary to ref.\cite{steinberg92}, the
dispersion is not cancelled since the two photons are not
frequency correlated.

We perform a Hong-Ou-Mandel (HOM) quantum interference experiment
to verify the indistinguishability of photons A and C
\cite{hong87}. Fig \ref{dip3x2km} shows the coincidence count rate
of the two detectors placed at the outputs of the BSM beam
splitter plus the laser clock ($C_1+C_2+t_0$) as a function of the
delay of Alice's photon, when the pump intensity is equal for the
two sources. The maximum visibility of the interference is limited
to $33\%$ in this case, due to the impossibility to discard the
(non interfering) events where two pairs are created in the same
source \cite{hdr03}. We observe a raw visibility of ($17\pm2\%$)
and a net visibility of ($28\pm2\%)$. Interestingly, this
interference occurs even if the photons are actually no more
Fourier Transform Limited but still indistinguishable when they
arrive at the beam splitter. The effective length of the wave
packets ($\approx$40 ps) is indeed much longer than the coherence
length given by the IF, due to chromatic dispersion in the fibers.
The observed width of the dip ($\approx 140 \mu m$) corresponds
roughly to the width expected from the IF, and not to the
effective length. An intuitive way to understand this is to note
that chromatic dispersion could be in principle compensated.
"Which path" information could then be gained, except within the
coherence length of the photons.

\begin{figure}[h]
\includegraphics[width=7 cm]{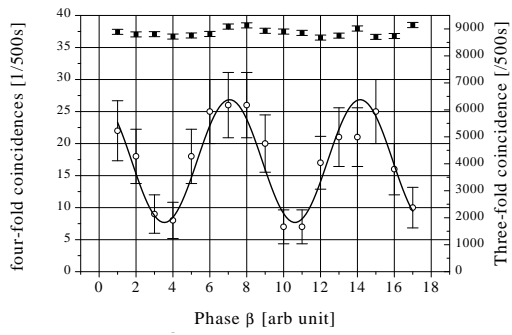}
\caption{ Teleportation of equatorial states. Open circles are raw
four fold coincidences and plain squares are three-fold
coincidences. The integration time for one point is 500 s.}
\label{telep6km55}
\end{figure}

To demonstrate the teleportation process, we teleport two classes
of qubit states: one class contains qubits that lie on the equator
of the Poincar\'e sphere (see Fig.2b, coherent superpositions of
$\left| 1,0 \right\rangle_{A}$ and $\left| 0,1 \right\rangle_{A}$
with equal amplitudes) and the other class contains the two poles
of the Poincar\'e sphere (the two time-bin themselves). For the
equatorial states, a successful teleportation implies Bob's photon
to be in a superposition state (see eq.\ref{teleport}). Thus, the
coincidence count rate of detector B, conditioned on a successful
BSM (4 fold coincidence $C_1+C_2+B+t_0$) is expected to vary as :
$ R_c=\frac{1+Vcos(\alpha+\beta)}{2} $, where V is the visibility
of the interference fringes, that can reach theoretically  the
value of one. From the measured visibility, we can compute the
fidelity for the equatorial states $F_{equator}=\frac{1+V}{2}$. If
there is no information about the BSM, Bob's photon will be in a
mixed state and no interference is expected. This can be
visualized by recording three fold coincidences ($C_1+B+t_0$).
\\
Fig. \ref{telep6km55} shows a typical teleportation result for
equatorial states. Four-fold and three-fold coincidence count
rates are plotted as a function of the phase $\beta$ of Bob's
interferometer. Four-fold coincidences are clearly oscillating and
a sinusoidal fit gives a raw visibility of $55 \pm 5 \%$, leading
to a fidelity of $F_{equator}=77 \pm 2.5\%$. Various measurements
for different values of $\alpha$ have been performed, leading to
similar fidelities.
\\
The analysis of the two other states represented by the north and
south poles of the Poincar\'e sphere is made by replacing the
interferometer by one fiber, by looking for detections at
appropriate times. The fidelity $F_{poles}$ is the probability of
detecting the right state when measuring in the north-south basis,
$F_{poles}=\frac{R_{correct}}{R_{correct}+R_{wrong}}$. The
measured fidelity for the $\left|1,0\right\rangle $ input state is
$(77\pm 3\%)$, and the measured fidelity for the$ \left|
0,1\right\rangle$input state is $(78\pm 3\%)$, leading to a mean
value of $F_{poles}=77.5\%$. Consequently, the total fidelity is
$\frac{1}{3}F_{poles}+\frac{2}{3}F_{equator}=(77.5\pm 3\%)$ higher
than the maximum value obtainable without entanglement (66.7
$\%$).\\ The fidelity depends both on the quality of the BSM and
on the visibility of the interferometers $V_{int}$, as :
\begin{equation}
F=V_{BSM}\cdot\frac{1+V_{int}}{2}+(1-V_{BSM})\frac{1}{2}
\end{equation}
where $V_{BSM}$ is the probability of a successful BSM. As the
visibility of the interferometers is close to one, the
teleportation fidelity is mainly limited by the quality of the
BSM, which could be improved by using narrower interferences
filters and by detecting the fourth photon as a trigger. This
would however result in lower count rates, and require an
improvement of the setup stability.

In conclusion, we reported a long distance quantum teleportation
experiment with a fiber delayed BSM, where Alice and Bob were
separated by 6.2 km of optical fibers and where the photon to be
detected by Bob covers an effective distance of 2.2 km. This
constitutes the first experimental realization of an elementary
quantum relay over significant distances. The main experimental
difficulty of such a configuration is to preserve the necessary
indistinguishability of the two photons involved in the BSM. For
an implementation outside the lab, the coherence length of these
photons should be dramatically increased, in order to tolerate
more fiber length fluctuations. This can be done, e.g. by placing
the non linear crystals in cavities or by using bright photon pair
sources \cite{tanzilli01} together with narrower filters.

The authors would like to thank Claudio Barreiro and Jean-Daniel
Gautier for technical support. Financial support by the Swiss NCCR
Quantum Photonics, and by the European project RamboQ is
acknowledged.

\end{document}